# Wave front sensor-less adaptive optics based on sharpness metrics for particle image velocimetry

M. Teich, J. Grottke, H. Radner, L. Büttner, and J.W. Czarske

*Technische Universität Dresden, Faculty of Electrical and Computer Engineering, Laboratory for Measurement and Sensor System Techniques, Helmholtzstrasse 18, 01062 Dresden, Germany*

*\*Corresponding author: martin.teich@tu-dresden.de*

**Optical distortions can significantly deteriorate the measurement accuracy in imaging systems. Such distortions can occur at fluctuating phase boundaries as well as multiple-phase flows and result from the accompanied refractive index changes. Due to multiple reflexes arising from a fluid flow setup, the usage of a wave front sensor (WFS) can be hindered. In this work we outline a wave front sensor-less approach which includes iterative aberration correction with a fast deformable mirror (DM). A combination of sharpness metric (SM) image evaluation and iterative optimization is demonstrated. The SM was measured for each image while adjusting seven Zernike modes (after Noll index enumeration) in their amplitude. The SM is used as an indicator for wave front aberrations without using a wave front sensor to correct wave front distortions that are generated by the DM. The proposed method allows for the reduction of systematic measurement uncertainties in fluid flow measurement techniques as particle image velocimetry (PIV). Five different sharpness metrics are demonstrated for reliable sharpness maximization with a deformable mirror of 69 elements. A systematic linear search (LS) algorithm was applied in order to find the optimal mirror shape. The reduction of measurement uncertainty for PIV measurements is shown. The iterative approach offers a way to reduce static or slowly changing wave front distortions in a fluid flow setup where a WFS is not applicable.**

## 1. INTRODUCTION

Adaptive optics has developed to an established method in many research fields of applied optics. Adaptive optics is used in astronomy [1] for compensating atmospheric turbulences. Another field of application is ophthalmology [2] in order to get sharp retina-images within the human eye or effectively perform vision correction. In microscopy [3] for biomedical applications, aberrations mainly result from different cell layers. Aberrations can also influence the point spread function (PSF) for high-resolution confocal microscopy [4]. The drawback of intrinsic aberration effects in a measurement system or from a measurement object can be compensated by applying spatial light modulators. Several methods exist to determine the wave front of a light wave. Beside interferometric methods and holography-based modal wave front sensing (HMWS) [5], the Hartmann-Shack wave front sensor [6] is a widely spread tool. Beside active in-situ correction of disturbed wave fronts, also post-processing approaches can be applied to improve the image quality, e.g. for removing turbulence effects [7] or distortion correction based on affine transformation algorithms. If the optical access for a wave front sensor is not available, i.e. a transmission measurement is not performable, then a sensor based closed-loop setup cannot be applied [8], except a Fresnel guide star approach is applicable [9], [10].

Wave front sensor-less adaptive optics is commonly used with a pinhole and a single photodetector [11], [12] allowing single light spot correction i.e. the intensity is maximized and the spot shape is improved (e.g. for confocal microscopy). There are also sensor-less concepts in terms of FFT-based aberration correction of low spatial frequencies [13] and by using the second moment of the image Fourier transform [14].

However, image based wide-field correction requires a CCD-camera and a more complex optimization strategy [15]. Usually an iterative approach is chosen, which does not require an additional sensor beside the imaging camera. Iterative methods are expected to be time-consuming due to the large parameter space. Either single elements of the actuator can be addressed or models for the expected optical distortion can be assumed. The latter one leads to a model based sensor-less correction [16] which often uses Zernike modes.

Particle imaging velocimetry (PIV) [17] is an image-based measurement technique that is used for fluid flow measurements. Seeding particles that follow the flow are added to the fluid. Usually a light-sheet illumination is generated, producing a homogenous light plane which is the measurement area from where the seeding particles scatter light to the detector (CCD camera). By evaluating the position shift of the cross-correlation (CC) peak between two consecutive frames, the particle movement and therefore the flow velocity can be determined [17]. The quality of the particle images is essential for low systematic uncertainties. The less sharp a particle image is, the broader is the cross-correlation peak and the larger the uncertainty in determining the spatial shift between two frames (see also Fig. 3).

In the present paper we show that it is possible to effectively reduce measurement uncertainties in PIV by using a sharpness metric (SM) and a systematic linear search (LS) algorithm which finds the optimal solution within several seconds by cycling through the deformable mirror modes. Similar approaches have been demonstrated in the past for static microscopy applications [18], but not for cross-correlation based flow measurements. In our work the source of aberration is the deformable mirror itself [16] since it gives a reproducible control on the experimental outcome in order to verify the best combination of sharpness metric and search algorithm for practical application.

The setup uses a large bandwidth actuator, namely a deformable mirror with a single-element settling rate of 1 kHz. For single-element testing of an 8-bit deformable mirror with 69 elements (DM), the parameter space has a size of $256^{69}$ which is far too large for a systemic search algorithm. Therefore we use Zernike mode testing [19] of seven modes after Noll index enumeration (from Z4 to Z10, see also Fig. 2). By applying 100 amplitude values for each mode it results in a much smaller search space of $100^7$ possible combinations. As a measure of image quality we tested five different sharpness metrics. Many sharpness metrics exist [20], [21]. It has to be noted that the sharpness metric has to fit to the observed scene and type of image [22].

Search algorithms can be divided into a systematic type as the linear search algorithm (LS) is, and non-systematic type as the Nelder-Mead simplex [23] or the stochastic gradient descent (SPGD) [18] are. All these algorithms show substantially different behavior in terms of the time needed to identify global extrema.

## 2. SETUP AND SOFTWARE

The experimental setup consists of a laser source (continuous wave, wavelength: 543 nm) which is combined with a cylindrical lens in order to generate a light sheet for illumination of the seeding particles (having a size of 10 µm in diameter) within a water basin. A microscope objective observes the light sheet plane from above. The image path is combined with a correction element namely a deformable membrane mirror (Alpao-DM69) with 69 elements. After reflection on the mirror membrane the light is focused onto a CCD camera (Basler Pilot piA640-210gm, 648x488 pixel, 210 Hz maximal frame-rate). The image data are evaluated on a workstation PC (CPU Intel Xenon CPU E5-1650 3.6 GHz, 12 Cores, GPU Nvidia Quadro K1200, 32 GB RAM, Win10 64bit) in order to determine the sharpness metric value in dependence on the Zernike-mode amplitude that was set to the deformable mirror. For data processing and sharpness metric evaluation MATLAB R2016a was used in combination with parts of the Image Acquisition and Optimization Toolbox. The membrane settling time is given by 10 ms whereas single element settling time is about 1 ms. The measured duration time values for sharpness metric calculation were 3 ms to 13 ms depending on the type of sharpness metric (see Table I). A time of 9 ms was needed for the acquisition of one PIV image (with GigE in MATLAB).

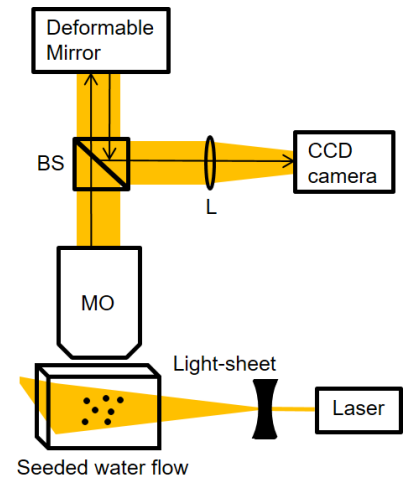

Fig. 1. Experimental setup for PIV measurements with a deformable mirror. Laser light (543 nm) is transformed to a light sheet by a cylindrical lens. Seeding particles within a water basin scatter light through a microscope objective (MO) to a deformable mirror (Alpao DM69). A beam splitter (BS) reflects the modulated light which is focused by a lens (L) to a CCD camera (Basler Pilot piA640-210gm).

## 3. METHODS

For PIV image distortion seven Zernike modes have been applied to the deformable mirror. The sharpness evaluation was performed with five different sharpness metrics under test regarding the calculation speed and the suitability for correct global extremum estimation (see also Table I). An open-source PIV algorithm was applied to the measured flow data [17] for three cases. The first one is the undisturbed case with a flat DM. The second case is the disturbed one where a set of Zernike mode amplitudes was loaded to the DM which deteriorate the image quality significantly. And the third case is the corrected one where a linear search (LS) algorithm was applied to systematically correct for the applied Zernike mode amplitudes.

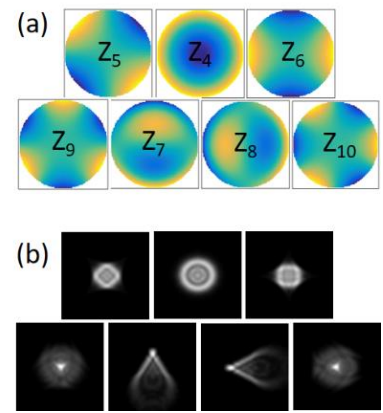

Fig. 2. (a) Seven Zernike modes generated with MATLAB and enumerated after Noll index. (b) Impact on the intensity distribution of the PSF that is comparable to an imaged seeding particle.

### A. Cross-correlation peak quality

In Fig. 2 is depicted the Zernike-mode enumeration after Noll index [24] that have been considered for image distortion and correction. Seven Zernike modes (from Z4 to Z10) have been considered for image correction. In Fig. 2(a) is shown the qualitative phase change within a circular pupil delivered by a deformable mirror. In Fig. 2(b) is shown the impact on a Gaussian light distribution, namely an imaged scattering particle. The intensity distribution of the PSF is modulated by the applied phase change. Defocus (Z4), astigmatism (Z5, Z6), coma (Z7, Z8) and trefoil (Z9, Z10) can be set to the deformable membrane mirror with RMS values from (-5 to +5) µm. This results in peak-to-valley values of about (25 to 30) µm depending on the mode (compare Fig. 2(a)).

Image distortions result in deteriorated cross-correlation functions, which lead to large systematic uncertainties for the velocity measurement. This behavior is demonstrated in Fig. 3. A sharp PIV image sequence (Ref) and disturbed image sequences ($Z_4$, $Z_7$) were evaluated concerning their cross-correlation (CC) peaks. At first the auto-correlation (AC) was calculated for each image. The position change of the cross-correlation peak with respect to the AC center gives the average movement of the seeding particles in the observation window (commonly known as interrogation window). For degraded images that are disturbed by aberrations (here shown for Z4, Z7), the position change measurement is worse since the cross-correlation peaks is broadened and deformed depending on the Zernike mode or a superposition of them.

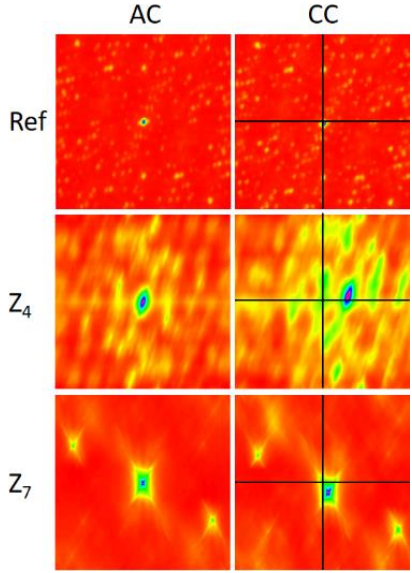

Fig. 3. Exemplary auto- (AC) and cross-correlation (CC) of consecutive images. By adding aberrations in form of Zernike mode $Z_4$ and $Z_7$ to the particle images the quality of the AC and CC peaks is degraded leading to a large systematic uncertainty for the velocity measurement.

### B. Sharpness metrics

Different types of sharpness metrics [21] have been applied in order to estimate sharpness and image quality of particle images. The following types of sharpness metrics have been used for feedback control of a deformable mirror. They are discussed in the following.

TABLE I
METRIC SINGLE IMAGE CALCULATION TIME

| Metric | Type | Calculation time |
|---|---|---|
| $S_1$ | Squared Laplace image | 3 ms |
| $S_2$ | Mendelsohn and Mayall | 3 ms |
| $S_3$ | Riemann tensor | 5 ms |
| $S_4$ | Steepness | 8 ms |
| $S_5$ | Maximal gradients | 13 ms |

The sum of the squared Laplace images $S_1$ is a metric that weights large grey-value differences in x and y direction. The image is convoluted with the Laplace-Kernel and the borders of the image are deleted. Here, M and N are the maximal available pixel number in x and y direction (M=648, N=488).

$$S_1 = \sum_{x=1}^{M} \sum_{y=1}^{N} \left( \frac{d^4 I_{x,y}}{dx^2 dy^2} \right)^2. \quad (1)$$

The histogram evaluation corresponds to a method by Mendelsohn und Mayall [25] and is a well-established approach. It is based on the fact that only a few grey-values exist for blurred images. Sharp images possess a large gray value distribution. The metric is a weighted sum of histogram intensities $H_k$ with a threshold T operating as a low intensity noise-filter.

$$S_2 = \sum_{k>T} k \cdot H_k. \quad (2)$$

The Riemann tensor metric $S_3$ [26] investigates the curve variation in Riemann space since the intensity distribution of an image can be mapped to Riemann space.

$$S_3 = \frac{1}{MN} \sum_{x=1}^{M} \sum_{y=1}^{N} | g_{x,y} |.$$
$$g_{x,y} = 1 + I_x^2 + I_y^2 - I_x^2 \cdot I_y^2. \quad (3)$$

The steepness metric $S_4$ was designed by us for the needs of the experiment. It is a sum of the quotient of maximal grey values $I_{max,Dk}$ for a particle and the value of its longest axis $D_k$. Note that certain Zernike mode distortions can lead to the deformation of particle images (e.g. elliptical shape for $Z_5$ and $Z_6$). This metric averages over all steepness values found in the image.

$$S_4 = \frac{1}{N} \sum_{k=1}^{N} g_k. \quad (4)$$

$$g_k = \frac{I_{\max, D_k}}{D_k}. \quad (5)$$

The metric of maximal gradients $S_5$ [21] evaluates the occurrence (frequency) of grey-value differences $d_i$ within the surrounding area of a particle (square of 21 by 21 pixel). The metric sums up all grey-value differences for the number of areas A around detected particles.

$$S_5 = \sum_{i=1}^{A} d_i. \quad (6)$$

In order to have a time-efficient feedback loop, the computational cost for sharpness metric calculation of each image is important. The computation time was measured with MATLAB Profiler (in ms) for the tested metrics. Note that the metric calculation time is not supposed to be much larger than the mirror membrane settling time (10 ms) and the image acquisition time (9 ms).

## 4. EXPERIMENTAL RESULTS

Several Zernike mode amplitude scans have been performed for different sharpness metric calculation methods. This has been done for PIV particle images during fluid flow. Fig. 4 shows the normalized metric behavior of five different metrics. The scans have been done in the available amplitude RMS range from (-5 to +5) µm. The behavior of all five metrics is symmetric around RMS = 0 µm. Note that the metrics possess different widths of decline, which mainly determines the number of necessary amplitude values that have to be set in order to find the extremum. The width of decline is therefore important for the overall correction speed in a feedback loop. Also notable is, that with higher Noll index of the Zernike mode the decline width increases significantly for all five metrics. Important is also that the reduction of noise is crucial for precise extremum estimation. Here a defocus ($Z_4$) of 2.7 µm was introduced to the measurement system by a centered lens with a long focal length of 140 mm. Metric $S_4$ accurately matches the defocus value during the scan whereas the other four metrics have a lower performance. By looking at the astigmatism behavior $Z_5$ and $Z_6$ (Fig. 4(b) and Fig. 4(c)) it turns out that only metric $S_4$ (steepness) performs well since the other metrics detect two symmetric $Z_5$ and $Z_6$ values as optimum which do not correspond to sharp images.

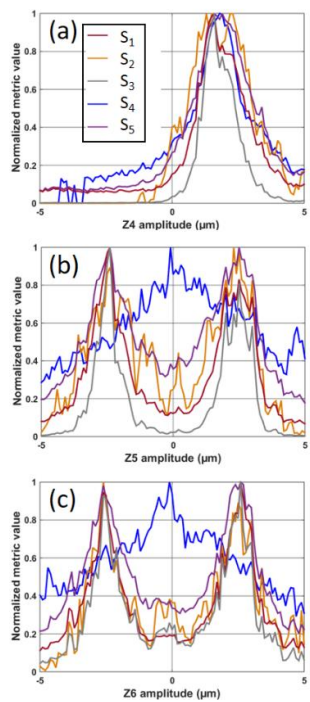

Fig. 4. Zernike mode amplitude scan of PIV images with introduced defocus by the deformable mirror. (a) The $Z_4$ scan reveals the correct $Z_4$ RMS value for a sharp image, namely +2.7 µm. (b) $Z_5$ scan. (c) $Z_6$ scan. Except $S_4$ (steepness) all other metrics exhibit a double-peak behavior whose maxima metric values do not correspond to sharp images.

A systematic optimization algorithm has been used, the linear search (LS). Linear search goes step by step through the parameter space and is testing one Zernike mode after another. The problem, that can arise, is, that the algorithm can be trapped in a local extremum for the applied metric. The advantage is, that it is relatively fast (a few seconds).

The correction method is currently not real-time capable for a free water surface in a basin, but it can be used for static aberration in an optical access or for slowly varying aberrations e.g. on a droplet or bubble surface.

In Fig. 5 is shown such a LS optimization procedure, which was performed during the laminar flow of the fluid in the water basin. The water surface was kept calm and was therefore not disturbing the measurement here. A set of randomly selected Z amplitudes from mode 4 to 10 was applied to the deformable mirror. The resulting PIV images appeared heavily blurred and deteriorated. The linear search needed 19.6 seconds to iteratively compensate the artificially introduced distortion. After correction the PIV images recovered in sharpness and the scattering particles appeared as symmetric Gaussian intensity distributions. Note that depending on the abort condition which mainly influences the time consumption, the recovered Z amplitude values after correction can vary by within 10 percent in comparison to the introduced distortion.

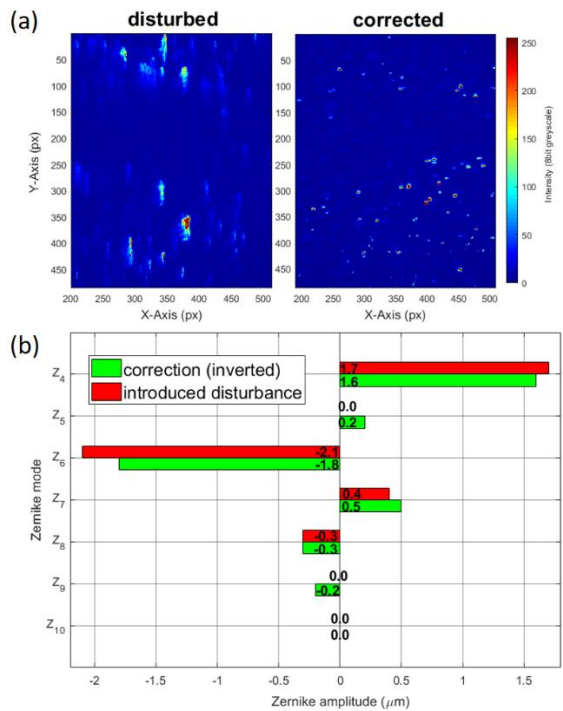

Fig. 5. (a) Iterative optimization during laminar flow of seeding particles in a water basin. The seeding particles appear blurred and deteriorated in shape for the disturbed case. (b) Applied (red) and iteratively determined (green) Zernike mode amplitudes of the deformable mirror.

Flow profile measurement have been performed in the water basin directly after a nozzle. A characteristic velocity field after the nozzle is formed. It was measured with PIV as it is depicted in Fig. 6a (undisturbed case). The flow direction is indicated here by the arrow located at the center of each interrogation window. The length of the arrows represents the velocity magnitude v. The peak velocity in the flow field had a value of 14 mm/s. By inserting aberrations in form of randomly generated Z amplitude values (compare Fig. 5), the PIV measurement is significantly deteriorated (disturbed). The flow profile cannot be resolved anymore and the peak velocity drops to 9 mm/s. After optimization the measurement quality has improved (corrected case) and the flow profile is measurable with the same accuracy as in the undisturbed case.

The relative standard deviation of the velocity magnitude $\sigma_v/v$ was calculated and is plotted in Fig. 6b. It is proportional to a system-intrinsic turbulence value arising from PIV evaluation. For very low velocity magnitudes in the shear layers $\sigma_v/v$ is high (up to 150%), whereas it is low for large velocity magnitudes (a few percent). In the disturbed case the measured velocity magnitude decreases and $\sigma_v/v$ increases in the corresponding interrogation windows (velocity arrows are overlaid here for better orientation in the flow field). A large $\sigma_v/v$ value leads to a lower detection rate of real turbulences in a measured flow field. After correction $\sigma_v/v$ is comparable to the undisturbed case over the whole field of view.

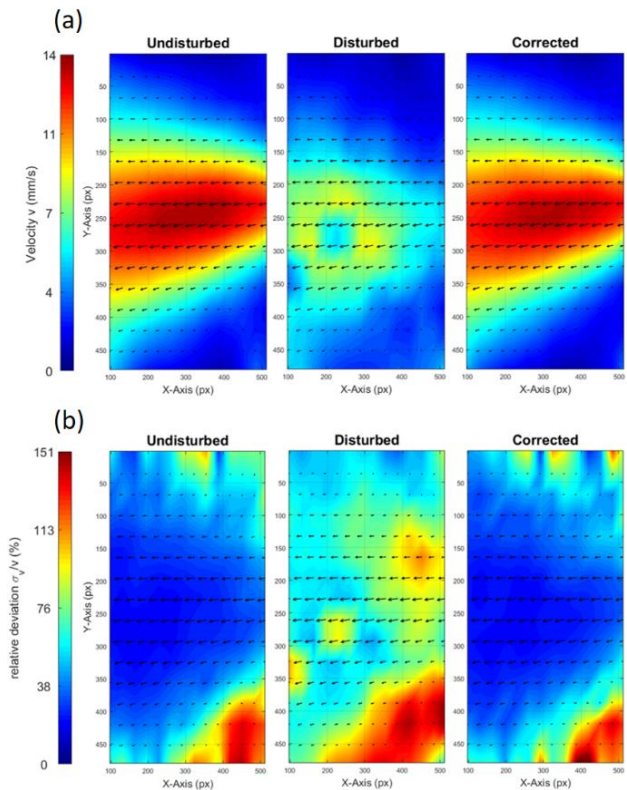

Fig. 6. (a) Flow velocity magnitude v after a nozzle. By inserting aberrations the measurement is significantly deteriorated (disturbed). After optimization the measurement quality has improved (corrected). (b) Calculated relative standard deviation $\sigma_v/v$ (flow field arrows are additionally overlaid for better orientation).

## 5. CONCLUSION AND OUTLOOK

In this paper we demonstrated an iterative approach for aberration correction i.e. without using a wave front sensor. The technique was tested with five different sharpness metrics. Seven Zernike modes have been combined in a superposition to generate an optical distortion. The applied sharpness metrics were characterized in terms of calculation time, decline broadening and the ability in correctly determining sharp images. The metrics of squared Laplace image (3 ms), Mendelsohn and Mayall (3 ms), Riemann tensor (5 ms), steepness (8 ms) and maximal gradients (13 ms) have been tested. Only the steepness metric ($S_4$) was able to determine sharp images for all seven Zernike modes.

A linear search algorithm (LS) was demonstrated as a reliable optimization method which requires about 20 seconds on a workstation PC for a complete optimization run.

We emphasize the importance of undisturbed particle images for a reliable PIV evaluation that is based on cross-correlation peak determination. For optical distortions that deteriorate the point-spread function of the system, the cross-correlation peak is broadened and deformed which leads to large systematic and statistic uncertainties in the velocity measurement.

Several attempts have been already undertaken in the past in order to reduce uncertainties in fluid flow measurements through an optical distortion [27] and even a fluctuating water surface [28]. In latter case a beam steering correction for a 0D measurement volume was accomplished. An interference pattern was optimized within the focal region in order to reduce measurement uncertainty. Smooth surface water waves are correctable due to their moderate wave height (200 µm) and frequency (300 Hz) that are within reach of common adaptive optics actuator amplitudes and frequency bandwidth. In future, surface waves can become correctable with low order Zernike modes for the improvement of PIV measurements through fluctuating air-liquid interfaces. Therefore the time budget of the correction system has to be reduced. A personal computer based operating system is not real-time capable. Therefore a field-programmable gate array (FPGA) could be of advantage in terms of image acquisition and metric calculation time.

**Funding Information.** Reinhart-Koselleck project (CZ 55/30) of the German Research Foundation (DFG).